# A Novel Temporal Attentive-Pooling based Convolutional Recurrent Architecture for Acoustic Signal Enhancement

Tassadaq Hussain, Wei-Chien Wang, Mandar Gogate, Kia Dashtipour, Yu Tsao, Xugang Lu,
Adeel Ahsan, and Amir Hussain

*Abstract*—In acoustic signal processing, the target signals usually carry semantic information, which is encoded in a hierarchal structure of short and long-term contexts. However, the background noise distorts these structures in a nonuniform way. The existing deep acoustic signal enhancement (ASE) architectures ignore this kind of local and global effect. To address this problem, we propose to integrate a novel temporal attentive-pooling (TAP) mechanism into a conventional convolutional recurrent neural network, termed as TAP-CRNN. The proposed approach considers both global and local attention for ASE tasks. Specifically, we first utilize a convolutional layer to extract local information of the acoustic signals and then a recurrent neural network (RNN) architecture is used to characterize temporal contextual information. Second, we exploit a novel attention mechanism to contextually process salient regions of the noisy signals. The proposed ASE system is evaluated using a benchmark infant cry dataset and compared with several well-known methods. It is shown that the TAP-CRNN can more effectively reduce noise components from infant cry signals in unseen background noises at challenging signal-to-noise levels.

*Impact Statement* — Recently proposed deep learning solutions have proven useful in overcoming some limitations of conventional acoustic signal enhancement (ASE) frameworks. However, the performance of these approaches in real acoustic conditions is not always satisfactory. In this study, we investigate the use of attention models for ASE. This is the first attempt that successfully employs a convolutional recurrent neural network (CRNN) with a temporal attentive pooling (TAP) algorithm for ASE. The proposed TAP-CRNN framework can be used to practically benefit the assistive communication technology industry, such as elderly people and school students with hearing loss. In addition, the derived algorithm can also benefit other signal processing applications, such as soundscape information retrieval, sound environment analysis in smart homes, physiological sound recognition, automatic speech/speaker and language recognition systems.

*Index Terms*— **Acoustic signal enhancement, convolutional neural networks, recurrent neural networks, temporal attentive-pooling.**

## I. INTRODUCTION

THE goal of acoustic signal enhancement (ASE) is to suppress the interfering noisy signal by minimizing unwanted distortions and modifying noisy input signals to desired clean ones. The acoustic signals are often distorted due to additive, convolution noises, or recording device constraints, which limit the performance of real-world applications, such as soundscape information retrieval systems [1-3], sound environment analysis in smart homes [4-6], physiological sound recognition [7-10], speaker recognition and verification [11-14], automatic speech recognition (ASR) [15-18], hearing aids [19, 20], and cochlear implants [21, 22]. Several ASE approaches have been proposed in the literature to alleviate the background noise problem. However, the performance of ASE in real-world acoustic environments remains a highly challenging task.

Traditionally, we assume that the target signals and noise follow specific distributions, and thus a gain function can be estimated to attenuate the noise components. Notable examples include the minimum-mean square error (MMSE) based algorithm [23-25], and Wiener filter [26]. For this class of approaches, a noise estimation approach is usually required to compute the statistics of noise signals. Well-known noise estimation approaches include minimum statistics (MS) [27], minima controlled recursive averaging (MCRA) [28], and improved MCRA [29].

Another group of traditional ASE algorithms include subspace-based techniques, that initially split a noisy acoustic signal into two subspaces: one for the clean acoustic signal and the other for noise components, and subsequently suppress the noise parts to reconstruct the clean acoustic signal. A notable subspace technique is based on singular value decomposition (SVD) [30].

The third group of traditional ASE approaches are model-based techniques; which derive a mathematical model based on human speech production and predict model parameters to perform noise reduction. Successful examples include the harmonic model [31, 32], linear prediction (LP) model [33-35], and hidden Markov model (HMM) [36, 37]. Later, matrix decomposition based methods like nonnegative matrix factorization (NMF) [38, 39] and sparse coding were proposed [40, 41] for ASE. These approaches develop dictionaries of the target signals and noises during the training stage. The noisy signals are decomposed in the online stage and clean signals can be reconstructed based on the dictionaries with corresponding activation metrics.

In recent years, deep learning (DL)-based approaches have emerged and shown great success for ASE. In a DL-based ASE system, a non-linear mapping function is estimated to transform noisy acoustic signals to clean acoustic signals. The system generally can be divided into three stages: feature extraction, non-linear mapping, and waveform synthesis. The feature extraction stage converts raw waveforms to acoustic features. Well-known acoustic features are spectral features that are obtained by applying short-time Fourier transform (STFT) on the raw-waveforms. Popular spectral features include Mel log-power spectrum (LPS) [42], log-power spectrum [43], and complex spectrum [44]. The non-linear mapping stage is formed by a neural-network model. In [45, 46] and [42, 47], fully connected neural networks and deep denoising auto-encoder (DDAE) are used as the mapping function. Meanwhile, convolutional neural network (CNN) [43], [48], [49], and long-short-term memory (LSTM) have also been used to form the mapping function. More recently, a weighted loss has been derived and used to train a gated recurrent unit (GRU) model for ASE [50]. Further, a combined CNN and RNN model architecture, termed convolutional recurrent neural network (CRNN), has shown near optimal results for ASE [51, 52]. Finally, a waveform synthesis based approach can also be used to generate waveforms from enhanced features. Specifically, if acoustic features are in the spectral domain, a waveform synthesis stage can adopt an inverse STFT (ISTFT) to convert the acoustic features to waveforms. Full end-to-end enhancement models have been recently derived in [48] to directly convert noisy waveforms to cleaner ones; in which case the feature extraction and waveform synthesis stages are omitted.

To increase real-world applicability, some approaches have been proposed to compress ASE models. In [53, 54], compressed DL models have been derived to meet low-latency and small model size requirements without compromising on the quality and intelligibility of the enhanced signal. In addition to single stage DL-based ASE systems, two-stage temporal convolutional modules (TCM) have been proposed by *Li et al.* [55] to deal with low signal-to-noise ratios (SNRs). In this system, a DL-based model first performs enhancement, and a secondary network then carries out post-processing to eliminate residual noises.

In this study, we extend the conventional CRNN framework by integrating a temporal attentive pooling (TAP) mechanism, the overall system termed TAP-CRNN, for ASE. The main idea of incorporating TAP in ASE is that local attention tries to capture short-term saliency features while global attention further emphasizes important features conditioned on the long-term context of the utterance. This innovative combination strategy is inspired by the intuition that a hierarchal structure of acoustic signals is encoded in the short and long-term context of acoustic signals.

In the proposed TAP-CRNN, the convolutional layers extract representative acoustic features and the RNN characterizes the long-short term temporal information. Further, the attention mechanism allocates significant segments in order to effectively train the enhancement model. The developed TAP-CRNN system has been evaluated for a benchmark infant cry enhancement task. The main contributions of our proposed TAP-CRNN system are:

(i) A new algorithm has been proposed i.e., the TAP algorithm, to provide an attention mechanism on top of a conventional CRNN framework for ASE;

(ii) The proposed TAP-CRNN aims to provide good discriminative power for target signal restoration and noise reduction compared to the CRNN and other state-of-the-art ASE models.

(iii) A challenging benchmark dataset is used to demonstrate superior performance of the TAP-CRNN in terms of four standardized metrics for the ASE task, including signal to interference ratio (SIR), signal to artifacts ratio (SAR), segmental SNR (SSNR), and signal distortion ratio (SDR). For comparison purposes, we tested performance of four widely used artificial neural network (NN) models, including DNN, CNN, RNN (long short-term memory (LSTM)), and CRNN.

The rest of this work is organized as follows. Section II reviews the related work. Section III discusses the proposed CRNN architecture and the TAP algorithm. The experimental setup and findings are presented in Section IV. Finally, Section V concludes the paper.

## II. RELATED WORK

Recent ASE research has mainly focused in three directions:

The first direction aims to use more suitable objective fictions to train the neural-network models. Traditionally L2 and L1 distance of the enhanced and clean reference signals are used as common objective functions. Since ASE is generally used as a pre-processor, when training the model, we consider the downstream task to achieve better performance. For speech signals as an example, it is more important to consider quality and intelligibility. Therefore, speech quality- or intelligibility-oriented objective functions can be designed. Along these directions, Fu et al., proposed to directly optimise the neural network model with a speech intelligibility-based objective function, namely short-time objective intelligibility (STOI) [48, 56]. In [57], a reinforcement learning approach is adopted to optimize the model parameters based on speech quality index, namely perceptual evaluation of speech quality (PESQ). Another approach derives an objective function that approximates the PESQ function to train the enhancement model [58]. More recently, MetricGAN [59, 60] was proposed to adopt a generative adversarial network to enforce the enhancement model to reach desirable metric scores. We also recently developed a novel STOI based deep neural network for multi-modal speech enhancement [90].

The second direction aims to incorporate complementary information from other modalities into ASE. Effective modalities include visual data [61-64][86-90], bone-conductive speech [65], and text information [66, 67].

The third direction seeks to explore better model architectures. Along this direction, successful approaches include: models with complex parameters [68-70], ensemble learning [71-73], dual path [74-76] and dual branch [77] architectures. In addition, attention mechanisms have also been adopted for ASE and shown excellent performance. In [78], the authors proposed a multi-task learning strategy along with multi-head self-attention mechanisms to analyse dependencies between the speech and noise signals. In [79], the authors investigated a multi-head attention network to estimate LPC for clean and noisy speech signals. In [80-83], the authors employed a convolutional neural network (CNN) and generative adversarial networks (GAN)-based architectures with attention mechanisms for end-to-end ASE.

combined input, i.e., CRNN output with attentive context, is further processed by the fully- connected layers to obtain the final enhanced acoustic LPS and noise LPS.

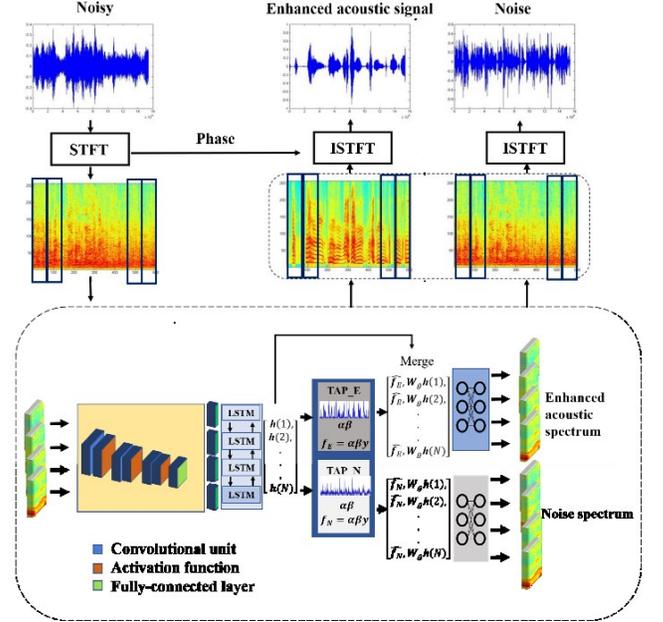

Fig. 2. The proposed TAP-CRNN-based ASE system.

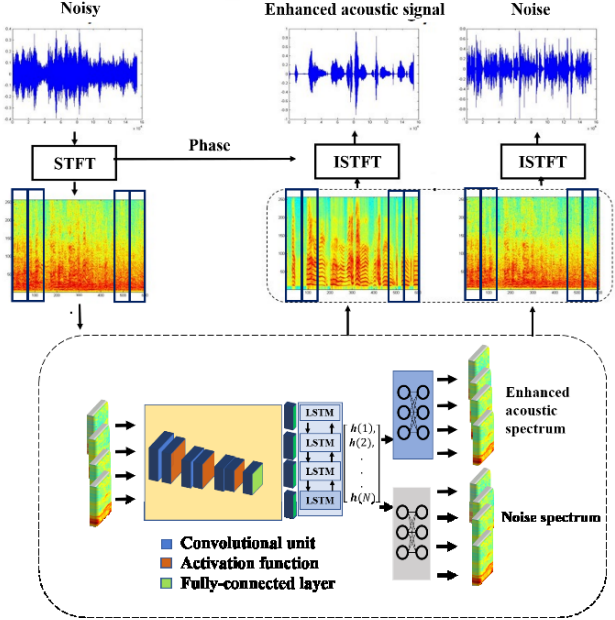

Fig. 1. CRNN-based ASE framework.

## III. THE PROPOSED TAP-CRNN BASED ASE SYSTEM

Figure 2 shows an overview of the proposed TAP-CRNN based ASE framework. When compared to the conventional CRNN, our proposed framework introduces an additional TAP mechanism.

The overall TAP-CRNN system first converts the time-domain noisy signals into LPS features. The LPS features are then processed by the ASE module to suppress noise components and accordingly enhance acoustic signals. Compared to the CRNN-based system, the TAP algorithm computes attentive contexts of the CRNN's output, which is then fed to the individual fully-connected layers to obtain enhanced acoustic features. Finally, the enhanced LPS features are converted back to the time domain using the ISTFT.

In the proposed model, the first part is the conventional CNN that takes the LPS as an input image to effectively preserve spatial information and reduce the number of parameters. Given a sequence of input vector $X = \{x(1),\dots,x(T)\}$ (with $T$ frames), the output of the CNN can be shown as $Y = \{y(1),\dots,y(T)\}$. Next, a bidirectional LSTM (BLSTM) efficiently explores both forward and backward temporal contextual information and extract global feature representation. The output of BLSTM is $\{h(1),\dots,h(T)\}$. During this stage, the TAP algorithm is applied to derive the attentive contexts. The attentive context is subsequently combined with CRNN's output for enhanced acoustic signal and noise, respectively. The

In the online stage, LPS features and phase components of the input noisy acoustic signals are first computed. The noisy LPS features are subsequently processed by the CRNN model. The output of CRNN is then processed by the TAP algorithm to obtain the attentive contexts of the enhanced spectra. The enhanced spectra is forwarded to the fully-connected layers to generate the enhanced LPS and noise LPS features. Finally, the enhanced LPS features are converted to the time domain using ISTFT along with the phase of the original noisy acoustic signals, as shown in Fig. 2. When training the model, we adopt the MSE loss to minimize the combination of the difference of estimated signal and reference signal and the difference of estimated noise and reference noise.

### A. Temporal Attentive Pooling (TAP) Algorithm

In this section, we describe the TAP mechanism in detail. The goal of TAP is to introduce attention blocks to the CRNN output by focusing on salient regions. To maximize the performance of TAP algorithms, we employed two complementary attention approaches i.e., local and global. The model focuses on attention equally in all regions as part of global attention. However, when generating the enhanced acoustic signal, the model only focuses on limited local regions. The purpose of global attention is to consider all CNN outputs as well as RNN's temporal summarization outputs. A previous study confirmed the benefits of local and global attention integration on a heart sound classification (regression) task [8]. In this study, we investigate for the first time, the effectiveness of integrating local and global attention on the challenging ASE task. The proposed TAP-CRNN model architecture illustrating integrations of the TAP mechanism is shown in Fig. 3. We use a simple concatenated layer to create the global attentive vector $c(t)$ for TAP by integrating information from CNN's output $y(t)$ and BLSTM's output as follows:

$$c(t) = \begin{bmatrix} W_c y(t) \\ W_r h(T) \end{bmatrix} \quad (1)$$

where $h(T)$ is the final hidden state of the temporal sequence, $W_c \in R^{N_c \times cnn_{dim}}$, and $W_r \in R^{N_r \times rnn_{dim}}$ are the parameter matrices used to concatenate $y(t)$ and $h(T)$, where $cnn_{dim}$ and $rnn_{dim}$ are the dimensions of the convolutional and LSTM layer, respectively. Next, $h(T)$ is combined with the LSTM's forward and backward last time step output.

The global attentive vector $c(t)$ is subsequently fed to the softmax layer to produce global attention weights $\alpha(t)$ (scalar) and is shown as:

$$\alpha(t) = softmax(u^T tanh(c(t) + b_g)) \quad (2)$$

where $u \in R^{(N_c + N_r) \times 1}$ is the vector to calculate the global attention weights shared by all time steps, and $b_g \in R^{(N_c + N_r) \times 1}$ is the global bias where $cnn_{dim}$ and $rnn_{dim}$ are the numbers of neurons for the last convolutional and last LSTM layer, respectively. The global attention weights are used to weight the local features from the CNN at each time step such as:

$$z(t) = \alpha(t) y(t) \quad (3)$$

Apart from global attention, we employed local attention to refine feature extraction and is calculated as:

$$\beta(t) = softmax(v^T tanh(W_l z(t) + b_l)), \quad (4)$$

where $W_l \in R^{N_l \times cnn_{dim}}$, $b_l \in R^{N_l \times 1}$ and $v \in R^{N_l \times 1}$ are the parametric matrices that are used to calculate local attention weight. These weights for local attention are used to weight features like:

$$f(t) = \alpha(t)\beta(t) y(t) \quad (5)$$

where $\beta(t)$ is the local attention output weight vector. The final attentive context is calculated as an average of the weighted outputs and is shown as:

$$\hat{f} = \frac{1}{T} \sum_{t=1}^{T} \alpha(t)\beta(t)y(t) \quad (6)$$

Next, we concatenate the attentive context $\hat{f}$ of the enhanced clean with the output of the CRNN $h(t)$ as the input $r(t)$ of the output layers such as:

$$r(t) = \begin{bmatrix} \hat{f} \\ W_g h(t) \end{bmatrix} \quad (7)$$

After that we obtain two $r(t)$ by TAP for separated noise and enhanced acoustic sound, respectively. The output layers are the fully-connected layers, where relationship between the input $r(t)$ and the output of the first hidden layer can be shown as:

$$a_1(t) = \sigma(W_1 r(t) + b_1) \quad (8)$$

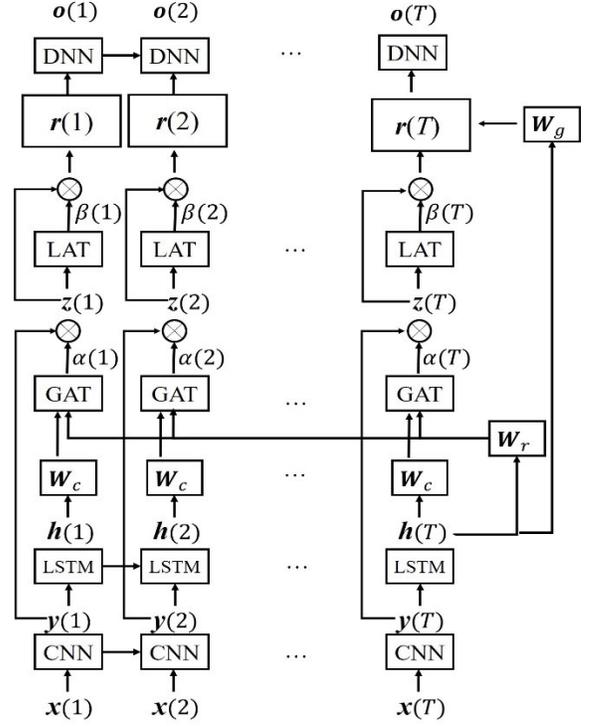

Fig. 3. The TAP_E (the same as TAP_N) of the TAP-CRNN model architecture.

where $\sigma(.)$ is the activation function, $W_1$ is the matrix, $b_1$ is the bias vector of the first hidden layer. Similarly, the relationship for the $q$-th hidden layer can be expressed as:

$$a_q(t) = \sigma(W_q a_{q-1}(t) + b_q), \; q = 2,...,Q, \quad (9)$$

where $Q$ represents the total number of neurons in the output layer. As a result, the link between the regression layer and the output layer can be summarized as follows:

$$o(t) = G(a_Q(t)) \quad (10)$$

where $G(.)$ is the linear function for the regression or output layer. and $o(t) \in R^{lps_{dim} \times 1}$, $t = 1,...,T$.

## IV. EXPERIMENTS

This section first presents the experimental setup including model training and evaluation details, followed by experimental results. In this study, we use infant cries as target signals from a benchmark dataset. Various noise types and SNR conditions are sampled to prepare training and testing sets. As noted in [84, 85], infant cries also possess short and long-term context structures, and the proposed TAP-CRNN can thus be suitably applied to enhance infant cry signals and compared with state-of-the-art models.

### A. Experimental Setup, Training and Evaluation

The experiments were performed using a benchmark infant cry dataset collected from 5 infants. For the training set, 400 infant cry utterances were randomly selected and corrupted with six noise types (Babble, Pink, a female speech, a male speech, a song, and cocktail party) at three SNR levels (-5dB, 0dB, and 5dB) to generate 400 (utterances) × 6 (noise types) × 3 (SNRs) = 7200 training utterances, and 100 infant cry utterances (different from those used in the training set) were selected for the test set. We created a mismatch condition test set by contaminating 100 clean utterances with two background noise signals (i.e., Fan and a song) at four mismatch SNR levels (i.e., 6dB, -2dB, 2dB, and 6dB) to generate 100 (utterances) × 2 (noise types) × 4 (SNRs) = 800 test utterances. In this work, 257 dimensional LPS features are extracted from the acoustic signal waveforms as acoustic features. Later, mean and variance normalization was applied to the input feature vectors to make the training process more stable.

To evaluate the performance of the proposed ASE system, we adopted two standardized evaluation metrics: segmental SNR (SSNR) and signal distortion ratio (SDR). The SSNR measures the average SNR values over short segments (15 to 20 ms) of acoustic signals. A higher value of SSNR indicates that the proposed TAP-CRNN suppressed the background noise effectively. The SDR measures the distortion of the enhanced signal where a lower SDR indicates less distortion in the enhanced signal. In contrast to standardized evaluation metrics where the enhanced acoustic signal is focused, the background noise signals also contain important information. Thus, we formulate the ASE as a monaural source separation task. More specifically, we intend to use the TAP algorithm to accurately separate the target and background noise signals. For evaluating such a task, we adopted two additional evaluation metrics, namely SIR and SAR, to estimate the performance.

For training the TAP-CRNN model, a noisy acoustic signal is given as input. Specifically, as detailed in Section III, the noisy and clean acoustic signals are first transformed into spectra via STFT for frame-based processing. Further, the LPS acoustic features of both clean and noisy spectra are subsequently processed by the ASE module (i.e., a deep neural network model). The goal of the ASE system is to minimize the reconstruction error between estimated and clean acoustic signals. In our work, 257 dimensional LPS is extracted from the acoustic signal waveforms as acoustic features. In the testing stage, the LPS features and the phase component of noisy acoustic input signals are first computed. The learnt parameters of the ASE modules then transform the noisy LPS features to obtain the enhanced LPS features. Finally, the phase of the original noisy acoustic signal is used and ISTFT applied to obtain the enhanced signal.

All ASE modules utilised to evaluate the performance were Tensorflow-based, with the RMSprop optimizer and a learning rate of 10e–5 and a batch size of 32 used to train the models over 300 iterations. The experiments were run on a single NVIDIA GeForce GTX 1080 Ti GPU with 11 GB of memory.

As compared to the original CRNN model, the TAP-CRNN has slightly increased parameter numbers: $W_c \in R^{N_c \times cnn_{dim}}$, $W_r \in R^{N_r \times rnn_{dim}}$, $W_g \in R^{N_g \times rnn_{dim}}$, $W_l \in R^{N_l \times cnn_{dim}}$, $b_g \in R^{(N_c+N_r) \times 1}$, $b_l \in R^{N_l \times 1}$, $u \in R^{(N_c+N_r) \times 1}$, and $v \in R^{N_l \times 1}$ for both TAP_E and TAP_N in the TAP-CRNN model. These additional parameters actually depend on the basic CRNN model.

### B. Evaluation of the TAP-CRNN

In this section, we first assess the performance of our proposed framework for unseen stationary and non-stationary noise types that are different from our training data. Table 1 shows the enhancement results for unseen stationary noise, namely fan noise. Note that with stationary noise, our model efficiently separated the spectral envelope of infant cry from the noise by preserving the enhanced acoustic speech signal. Overall, the proposed framework not only demonstrated better performance for high SNRs but also proved its effectiveness in separating infant cry spectral envelope from the noise envelope even at low SNRs (i.e., -2 dB and -6 dB). Table 2 presents the average SDI, SIR, and SAR performance for the TAP-CRNN framework on the other two unseen non-stationary noises, namely a *song* and a *female speech*. Despite a clear improvement in the scores, the performance of TAP-CRNN for stationary noise (Table 1) is better compared to non-stationary noise conditions (Table 2). The results in Table 1 and Table 2 demonstrate the ability of TAP-CRNN to exploit spatio-temporal information in an efficient manner by exhibiting better results for both unseen stationary and non-stationary noise conditions. Further, in addition to SDR, SIR, and SAR, Tables 1 and 2 display the SSNR score yielded by the TAP-CRNN framework under stationary and non-stationary noise conditions. From the results reported in the Tables, we can note that TAP-CRNN demonstrated better SSNR performance for unseen stationary noise conditions compared to non-stationary noise conditions at different SNR levels.

TABLE.1. EVALUATION OF THE TAP-CRNN FOR A STATIONARY NOISE TYPE (FAN).

|  | SDR | SIR | SAR | SSNR |
|---|---|---|---|---|
| -6dB | 3.653 | 21.395 | 3.762 | 13.375 |
| -2dB | 4.158 | 23.358 | 4.231 | 13.797 |
| 2dB | 4.421 | 24.696 | 4.478 | 14.595 |
| 6dB | 4.803 | 26.479 | 4.843 | 15.162 |

TABLE.2. EVALUATION OF THE TAP-CRNN FOR TWO NON-STATIONARY NOISE TYPES (SONG AND FEMALE SPEECH)

|  | SDR | SIR | SAR | SSNR |
|---|---|---|---|---|
| -6dB | 4.774 | 18.234 | 5.225 | 12.120 |
| -2dB | 5.572 | 20.655 | 5.894 | 12.597 |
| 2dB | 5.996 | 22.824 | 6.190 | 13.769 |
| 6dB | 6.206 | 24.614 | 6.323 | 13.893 |

## C. Comparison with Different Neural Network Architectures

We compared the performance of the TAP-CRNN framework against four well-known DL-based ASE frameworks, including DNN, CNN, RNN, and CRNN, selected as baselines. In the DNN, we used six fully connected layers each having 512 units. The CNN model had two convolutional layers; each having 32 filters with a kernel size of 3×1 and 2×1, and 3 fully-connected layers containing 512 units in each layer. For the RNN, two bidirectional LSTM layers each with 256 units, followed by two fully-connected RNN layers each having 256 neurons are used. For the abovementioned baselines, the exponential linear unit (ELU) function was employed as an activation function. In the CRNN, the input noisy LPS features are first processed by the CNN at each time stamp and subsequently fed to the two layers of bidirectional LSTM, each having 128 units. The output of the bidirectional LSTM is later concatenated and processed by the two independent DNNs, each having 2 hidden layers with 128 units, respectively.

Similar to CRNN, the TAP-CRNN has additional TAP blocks as described in section 1.2. Each block of TAP consists of an individual DNN, which has two feedforward layers of size 256 units. One of two subsequent blocks of TAP is used for ASE, where the enhanced acoustic signal is attained by the DNN, the other TAP block recovers the noise signal. The overall architecture of TAP-CRNN is illustrated in Fig. 2. The output of the subsequent layers of CRNN is processed by the two independent TAP blocks (DNNs) to fully reconstruct the enhanced acoustic signal by separating the noise signal. The activation function used for CRNN and TAP-CRNN is tangent hyperbolic.

Figure 5 shows a summary of average SDR, SIR, SAR, and SSNR scores yielded by the TAP-CRNN against four baseline systems at different SNR levels under stationary and non-stationary noise conditions. For all performance evaluation metrics, a high score describes better performance in terms of noise reduction (SSNR) and noise separation (SDR, SIR, and SAR). Overall, an improvement in performance can be seen for all frameworks from -6 dB SNR to 6 dB SNR. However, the TAP-CRNN model performed exceptionally well over the baseline systems even under low SNR levels. By looking at Fig. 4, we can observe that both DNN and RNN exhibited almost similar behavior for SDR, SRI, and SAR metrics at -6 dB SNR level. However, the RNN maintained a better SSNR score compared to DNN at -6 dB SNR level. Despite the efficient performance achieved by the DNN to recover the enhanced acoustic signal, we observed that if temporal information is added to the DNN framework through a bidirectional LSTM, the framework can achieve much better results by appropriately controlling the stationary and non-stationary noise distortions. CNNs proved to be effective at imitating spatial context for acoustic speech enhancement. The results of CNN demonstrated better performance compared with bidirectional RNN and DNN under mismatch testing conditions. It exhibited a similar and even slightly better performance for SSNR than the TAP-CRNN at high SNR conditions (i.e., 2 dB and 6 dB). However, it failed to show similar behaviour under low SNR conditions. Since CNN and RNN both efficiently recovered the enhanced acoustic speech signal, it was expected that the merger of CNN and RNN with bidirectional LSTM layers (known as CRNN), would further enhance the system capability and generalization performance. However, the results demonstrated a blended response by maintaining good performance for SDR and SIR at -6 dB, -2 dB, and 2 dB SNRs, but failed to perform well for SSNR and SAR. This could be on account of the CNN model learning feedforward spatial information while dealing with stationary and non-stationary distortions and forwarding it to the bidirectional LSTM which deals with both forward and backward temporal information. The output of bidirectional LSTM turned out to be a mixed representation learned by the bidirectional LSTM (i.e. spatial feedforward information with backward and forward temporal information), resulting in blended performance, as shown in Fig. 4.

To dissociate the blended information learned by the CRNN, a TAP mechanism is adopted to help the CRNN focus on the significant frame by adopting global and local attention weights and deciding where to focus attention when generating the enhanced acoustic signal or noise. The results of TAP-CRNN demonstrated excellent performance when compared with the baseline systems for all performance evaluation metrics except for SSNR at SNR = 6 dB, where CNN showed greater improvement compared to the TAP-CRN. Figure 5 further illustrates the comparative performance of the TAP-CRNN by displaying the visualization of attention weights, which show that the addition of a TAP mechanism to the CRNN architecture can yield performance improvement and provide better acoustic signal enhancement.

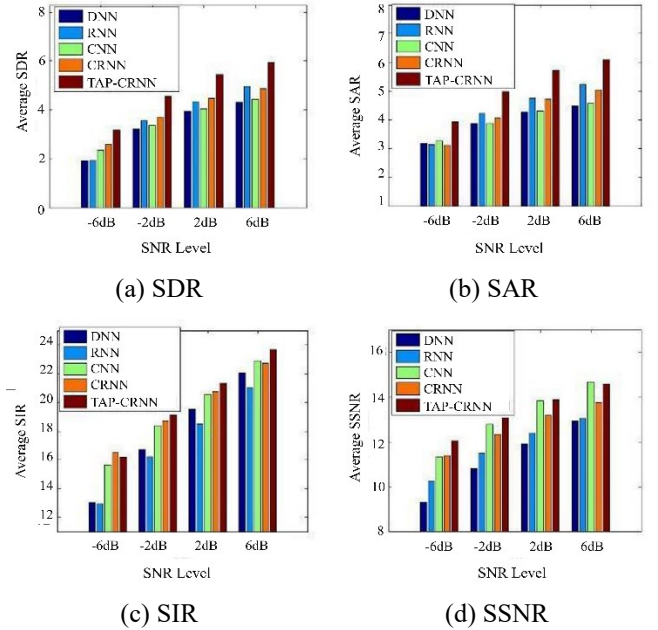

(a) SDR  (b) SAR
(c) SIR  (d) SSNR

Fig. 4. Evaluation results of different neural network models on the benchmark infant cry dataset.

## D. Spectrogram Analysis and Comparison

To visually compare the ASE performance of TAP-CRNN and deep learning models, we next present the spectrograms of the enhanced acoustic speech signals. Fig. 5 displays the spectrogram of enhanced acoustic signals achieved by DNN, RNN, CNN, CRNN, and TAP-CRNN models. Fig. 5(a) and (b) shows the spectrogram of noisy acoustic signal (termed as Noisy) corrupted with non-stationary noise (i.e., *male speech*) at -6 dB SNR and clean infant cry acoustic signal (termed as Clean) for comparison. By looking at Fig. 5, it is apparent that all methods successfully suppressed the background noise components while dealing with non-stationary noise and effectively restored the low frequency acoustic regions at very low SNR level (-6 dB) as shown in rectangular boxes. From Fig. 5(f), it is worth noting that CRNN effectively removed the background noise from the noisy signal, however, it misjudged the region in the rectangular box as an acoustic region which shows that it has learned blended information. However, the TAP-CRNN restored low frequency acoustic regions shown in the rectangular box closer to the clean utterance spectrogram by helping the CRNN focus on significant frames using global and local attention weights when generating the enhanced acoustic signal.

To further investigate the effect of the TAP mechanism, we interpret the weights of the sequence generated by CRNN and TAP-CRNN in Fig. 6. The displayed visualization corresponds to a randomly selected utterance from the infant cry dataset contaminated with unseen non-stationary noise. As can be seen in Fig. 6, the TAP learns alignments that correspond very strongly with target acoustic signals by focusing and giving equal attention/importance to the significant frames. The attention mechanism has correctly identified and restored the frames by discriminating noise and acoustic signal components by assigning high weights to the salient regions or frames in noisy signals.

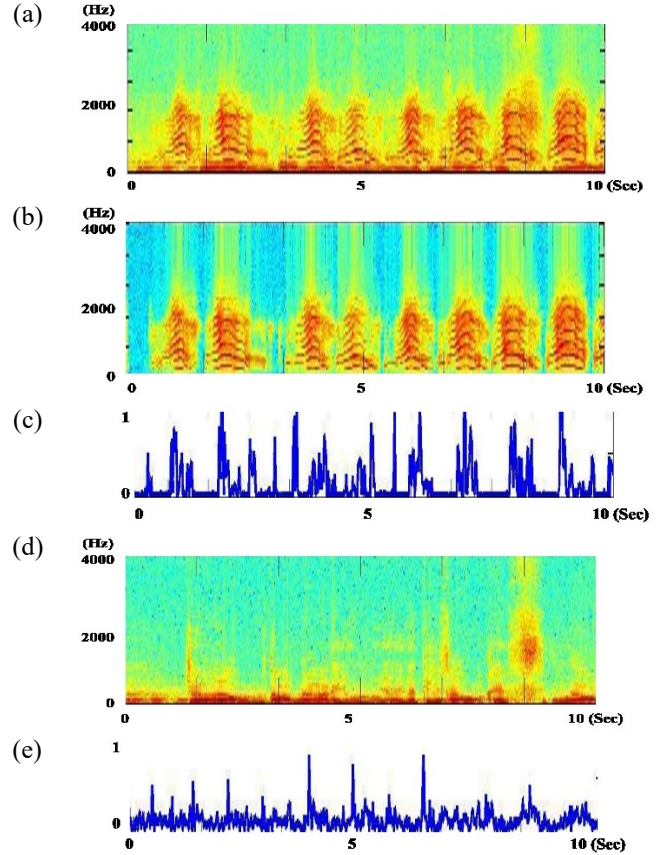

Fig. 6. (a) Spectrogram of a noisy signal; (b) Spectrogram of the enhanced acoustic speech signal; (c) Attention weights of TAP_E; (d) Spectrogram of the noise signal; (e) Attention weights of TAP_N.

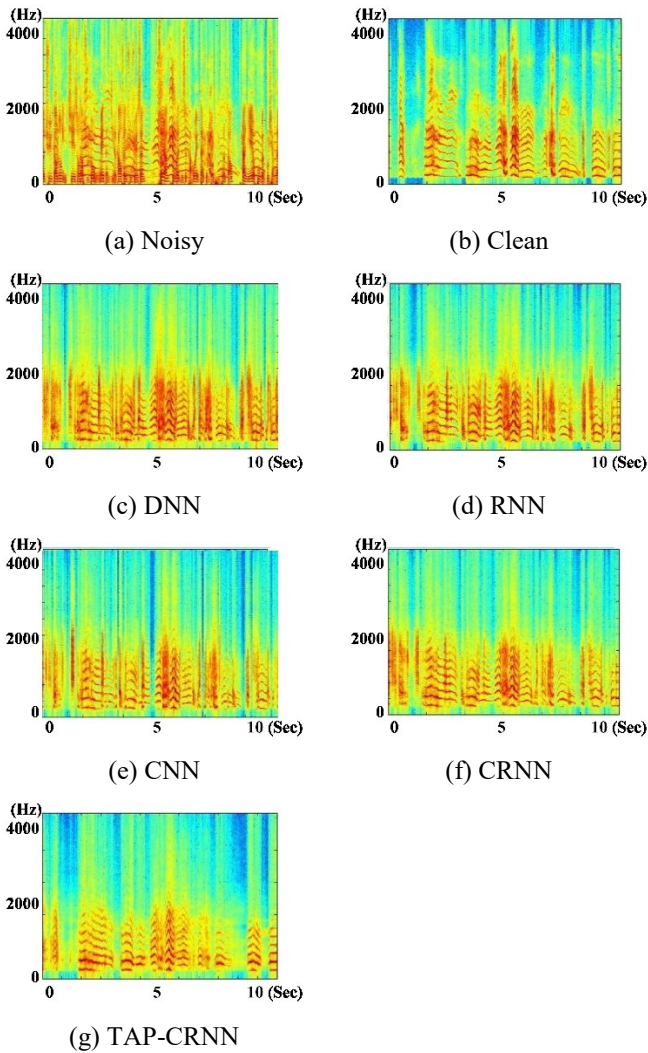

Fig. 5. Attention weights visualisation using the TAP mechanism.

## V. CONCLUSIONS

In this study, we investigated the novel use of attention models for ASE. Experimental results confirm the integration of a state-of-the-art CRNN model with the TAP mechanism delivers enhanced ASE performance compared to popular deep learning-based approaches. In particular, the results demonstrate the unequal importance of segments of each frame for ASE, which motivates the development of our proposed TAP mechanism that considers both local and global regions of the sequence and extracts significant features of regions of the sequence along with significant features of each frame. The global attention mechanism identifies salient parts of the sequence whereas the local attention mechanism minimizes the attention error. Comparative experiments show that the TAP-CRNN possesses robust discriminative power for target signal restoration and noise reduction compared to a conventional CRNN model. The effectiveness of the proposed TAP-CRNN mechanism is also demonstrated under mismatched non-stationary environments at severe SNRs in relation to other well-known deep learning based ASE frameworks.

It should be noted that the performance of our proposed TAP-CRNN has been analyzed (trained and tested) using a relatively limited infant cry dataset and formal subjective evaluation has yet to be carried out. In future work, we aim to consider more diverse training data and different types of non-stationary noises using a range of benchmark datasets to further scrutinize the performance of TAP-CRNN, in relation to more recent state-of-the-art methods. Further, we will explore noise-aware and SNR-aware training for TAP-CRNN in an attempt to further enhance system performance in real time environments. There is also a need to identify an approach to optimally determine the size of truncated segments according to application constraints, noise types and SNR levels. Also note that in the present pilot study, our aim was to prove the effectiveness of incorporating two types of attention mechanism models. A detailed theoretical analysis and performance complexity trade-offs are now required to be carried out in order to define the attainable performance of our TAP-CRNN model.

Finally, we will build on our recent works [86-90] and extend the TAP-CRNN model for multi-modal speech enhancement and exploit intelligibility-oriented loss functions for training. We will also investigate compressed versions of TAP-CRNN to contextually exploit multi-modal information (such as audio-visual lip-reading), with aims to meet strict latency (e.g. <10ms) constraints and generalized performance requirements for next-generation multi-modal hearing aids and listening devices. Applications where ideal real-time conditions are not required will also be explored, such as teleconferencing, soundscape information retrieval, physiological sound recognition, and entertainment (e.g. speech enabled AR/VR) systems.

## ACKNOWLEDGEMENTS

The authors are very grateful to the anonymous reviewers for their insightful comments and suggestions which have helped improve the quality of this paper. A. Hussain, A. Ahsan, M. Gogate and T. Hussain are supported by the UK Engineering and Physical Sciences Research Council (EPSRC), grant reference EP/T021063/1.